\documentclass[10pt,preprint]{emulateapj}
\usepackage{url}
\begin{document}

\shorttitle{} %

\shortauthors{Shen et al.}

\title{Fine Magnetic Structure and Origin of Counter-Streaming Mass Flows in a Quiescent Solar Prominence}

\author{Yuandeng Shen\altaffilmark{1,2,3,4}, Yu Liu\altaffilmark{1}, Ying D. Liu\altaffilmark{2}, P. F. Chen\altaffilmark{4}, Jiangtao Su\altaffilmark{3}, Zhi Xu\altaffilmark{1}, \& Zhong Liu\altaffilmark{1}}

\altaffiltext{1}{Yunnan Observatories, Chinese Academy of Sciences, Kunming 650216, China; ydshen@ynao.ac.cn}
\altaffiltext{2}{State Key Laboratory of Space Weather, Chinese Academy of Sciences, Beijing 100190, China}
\altaffiltext{3}{Key Laboratory of Solar Activity, National Astronomical Observatories, Chinese Academy of Sciences, Beijing 100012, China}
\altaffiltext{4}{School of Astronomy \& Space Science, Nanjing University, Nanjing 210023, China}

\begin{abstract}
We present high-resolution observations of a quiescent solar prominence which was consisted of a vertical and a horizontal foot encircled by an overlying spine, and counter-streaming mass flows were ubiquitous in the prominence. While the horizontal foot and the spine were connecting to the solar surface, the vertical foot was suspended above the solar surface and supported by a semicircular bubble structure. The bubble first collapsed and then reformed at a similar height, finally, it started to oscillate for a long time. We find that the collapsing and oscillation of the bubble boundary were tightly associated with a flare-like feature located at the bottom of the bubble. Based on the observational results, we propose that the prominence should be composed of an overlying horizontal spine encircling a low-lying horizontal and a vertical foot, in which the horizontal foot was consisted of shorter field lines running partially along the spine and with the both ends connecting to the solar surface, while the vertical foot was consisted of piling-up dips due to the sagging of the spine fields and supported by a bipolar magnetic system formed by parasitic polarities (i.e., the bubble). The upflows in the vertical foot were possibly caused by the magnetic reconnection at the separator between the bubble and the overlying dips, which intruded into the persistent downflow field and formed the picture of counter-streaming mass flows. In addition, the counter-streaming flows in the horizontal foot were possibly caused by the imbalanced pressure at the both ends.
\end{abstract}


\section{INTRODUCTION}
Solar prominences are cloud-like cool and dense plasma suspended in the extremely hot, tenuous corona, and prominence plasma is typically one hundred times cooler and denser than the surrounding coronal plasma\citep{mackay10}. The eruptions of prominences are vitally important for space weather forecast since they usually lead to coronal mass ejections that can significantly affect the interplanetary space\citep{liuy14,schmieder15}. Previous observations have indicated that a prominence is consisted of a main spine and a few lateral feet (or barbs)\citep{martin98}, and all these structures are composed of many field-aligned thin threads that carry intriguing counter-streaming mass flows\citep{zirker98,lin03,lin05,schmieder08,berger11,chen14,yang14,yan15}. So far, it has been widely accepted that prominences are supported by solar magnetic fields\citep{tandberg95}, and theoretical works have been performed to understand the prominence physics\citep{kippenhahn57,kuperus74,martin94,aulanier02,amari03}. However, the basic magnetic structure and the origin of counter-streaming mass flows in prominences are hitherto unclear, although these issues are crucial for understanding the prominence physics as well as the initiation of coronal mass ejections\citep{shibata11}.

Currently, there are three main discrepancies about the magnetic structure and counter-streaming mass flows in prominences. The first is about the magnetic field distribution in prominences. Some high-resolution observations showed that prominences are consisted of horizontal thin threads\citep[e.g.,][]{casini03,okamoto07,heinzel08,ahn10}, while others indicated that prominences are consisted of vertical thin threads\citep[e.g.,][]{berger08,chae08}. Recently, \cite{dudik12} argued that this discrepancy is largely depending on the observing angle with respect to the prominence axis. The second is about the relationship between the ends of prominence feet and the photospheric magnetic fields. Some observations showed that prominence feet are fixed at parasitic (minority) polarities that are opposite in polarity to the surrounding network elements\citep[e.g.,][]{martin94}, but other studies indicated that they are possibly terminated around small-scale polarity inversion lines of the photospheric magnetic fields\citep[e.g.,][]{wang01,aulanier98, aulanier98b,chae05,li13}. It should be noted that in \cite{martin94} prominence feet are composed of magnetic field lines running off the prominence spine and rooted in parasitic polarities, while \cite{aulanier98} and \cite{aulanier98b} suggested that prominence feet are composed of magnetic dips overlying the secondary photospheric inversion lines formed by parasitic polarities. The third is about the origin of counter-streaming mass flows in prominences. Some studies suggested that counter-streaming flows are the manifestation of longitudinal oscillations of thin threads\citep[e.g.,][]{lin03,chen14}, while others proposed that they are caused by force non-equilibrium at the two ends of threads\citep[e.g.,][]{low05,chen14}. These discrepancies about prominence have puzzled soar physicists many years. To reconcile these discrepancies, high-resolution and multi-wavelength observations are necessary, as well as the measuring method of magnetic spectropolarimetry.

In this letter, we present a quiescent prominence recorded by the New Vacuum Solar Telescope\citep[NVST;][]{liuz14}, which showed both of horizontal and vertical structures and intriguing counter-streaming mass flows. In addition, a sharply defined bubble structure is observed below the vertical foot of the prominence, which showed interesting collapsing, reformation, and oscillation dynamics. Thanks to the high temporal and spatial resolution observations of the NVST, we are able to study the fine magnetic structure and the origin of the counter-streaming mass flows in the prominence, as well as the physics of the dynamical bubble structure. Instruments and observations are described in Section 2, main observational results are presented in Section 3, and interpretation and discussions are given in Section 4.

\section{INSTRUMENTS AND OBSERVATIONS}
The NVST is an one-meter ground-based solar telescope operated by Yunnan Observatories of Chinese Academy of Sciences. Due to the influence of the Earth's turbulent atmosphere, the raw solar images taken by the NVST are reconstructed using high-resolution imaging algorithms\citep[see,][]{liuz14}. The reconstructed NVST H$\alpha$ data has a cadence and a spatial resolution of 12 seconds and 236 km, respectively. The extreme-ultraviolet (EUV) images taken by the Atmospheric Imaging Assembly \citep[AIA;][]{lemen12} onboard the {\it Solar Dynamics Observatory} \citep[{\it SDO};][]{pesnell12} are used in the paper, whose cadence and spatial resolution are 12 seconds and 880 km, respectively. In addition, full-disk H$\alpha$ images taken by the Kanzelh\"{o}he Solar Observatory (KSO) H$\alpha$ telescope and magnetograms taken by the Helioseismic and Magnetic Imager \citep[HMI;][]{schou12} onboard the {\sl SDO} are also used to aid our analysis.

\begin{figure*}[thbp]
\epsscale{0.9}
\plotone{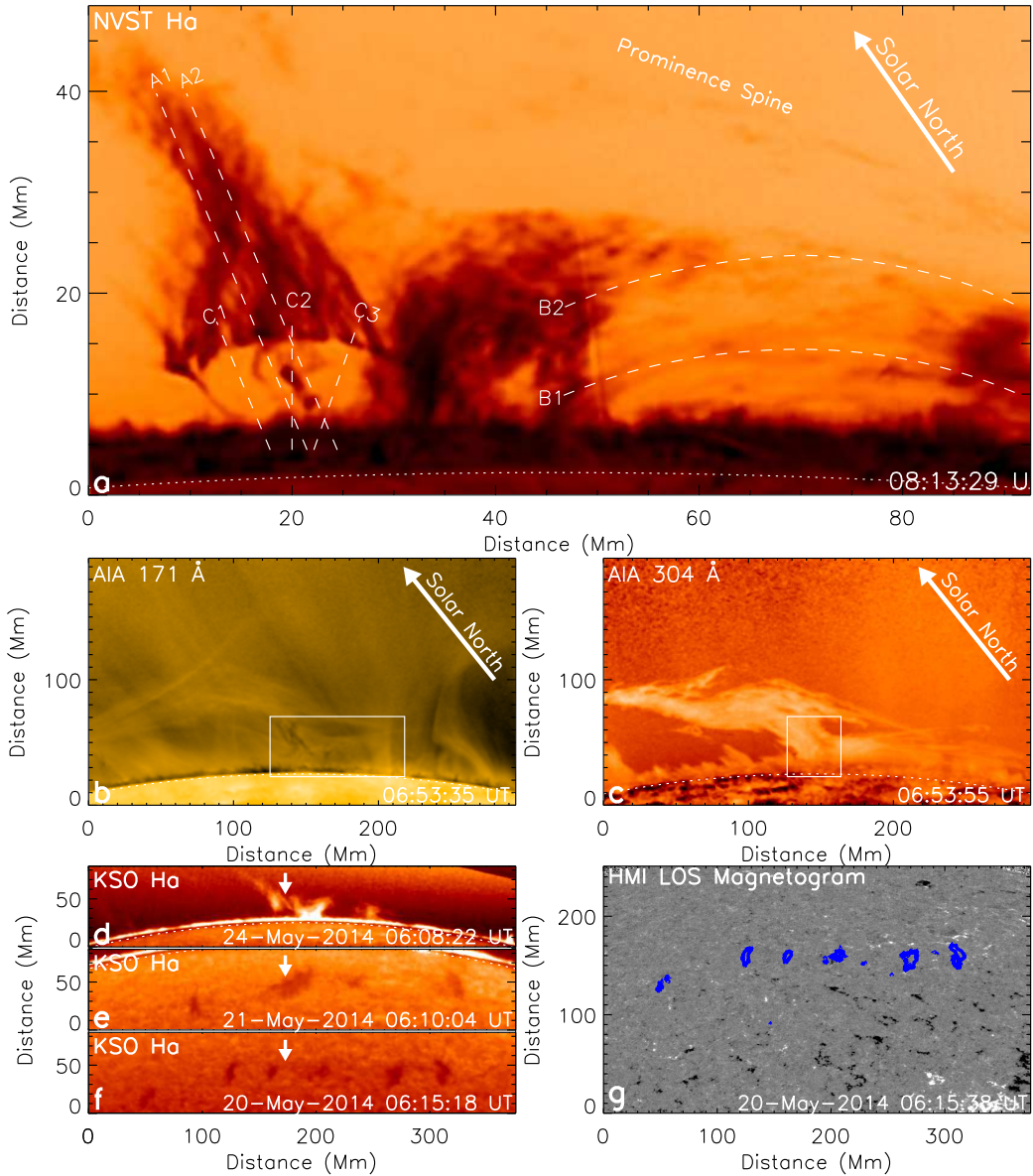}
\caption{Multi-wavelength observations show the quiescent prominence. a,  NVST H$\alpha$ negative image. The dashed lines A1--A2, B1--B2 and C1--C3 are used to obtain the time-distance diagrams. b and c, AIA 171 \AA\ image. The white box shows the field-of-view of panel a. c, AIA 304 \AA\ image. The white box shows the field-of-view of the panels in Fig. 2. The white arrows in panels a--c point to the solar north. d--f, KSO H$\alpha$ images. The white arrows indicate the prominence foot containing the bubble. g, HMI LOS magnetogram overlaid with the prominence profile determined from the KSO H$\alpha$ image on 20 May 2014 at 06:15:18 UT. See animations 1 and 2 for the all NVST and AIA images. 
\label{fig1}}
\end{figure*}

\begin{figure*}[thbp]
\epsscale{0.9}
\plotone{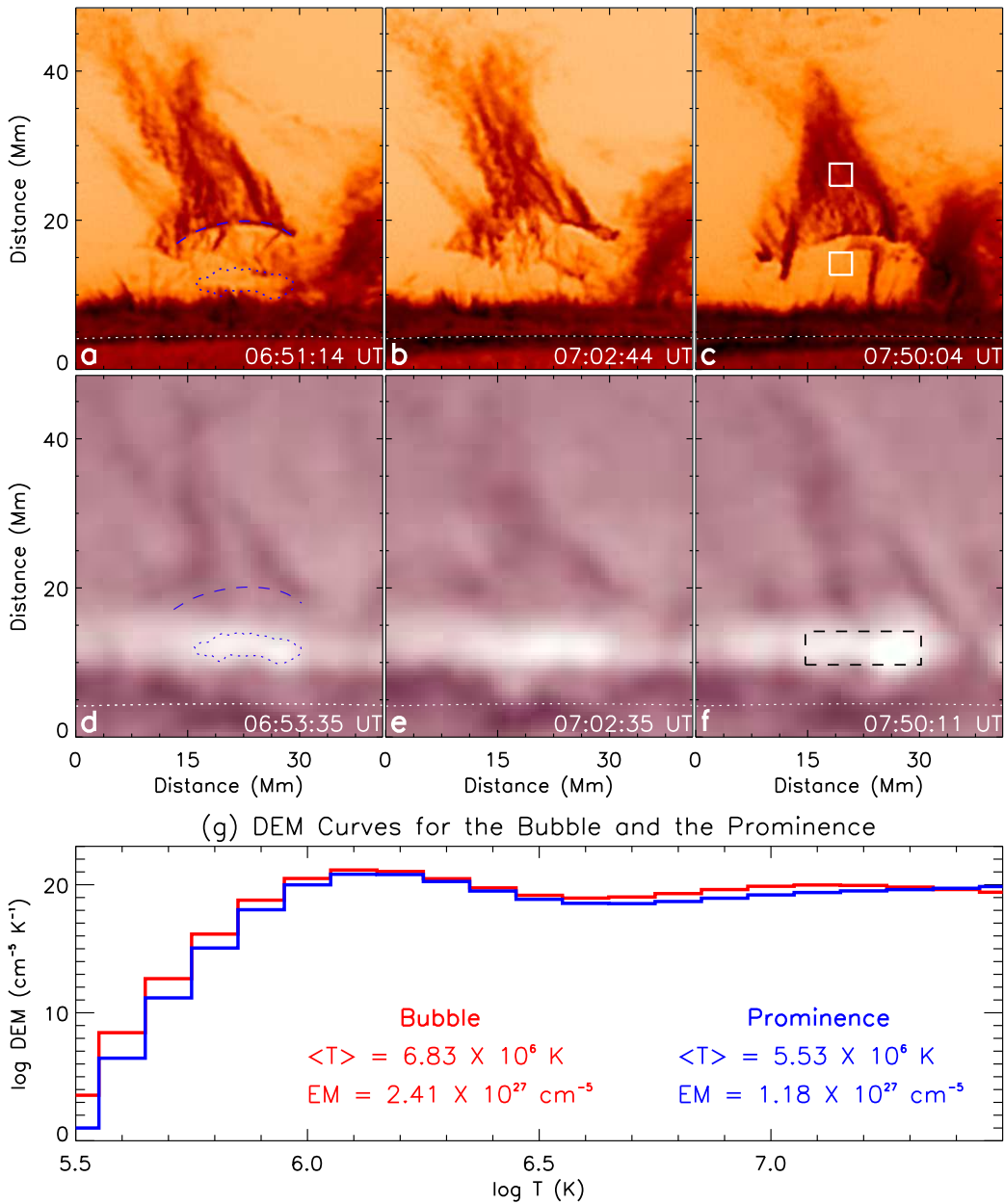}
\caption{Spatial and temperature evolution of the bubble and the vertical prominence foot. a--c, NVST H$\alpha$ negative images. d--f, AIA 211 \AA\ images. g, DEM curves derived from the box regions (ref. panel c) over the bubble and the prominence by using the six hot channels of AIA. The box in panel f shows the region used to measure the brightness shown in Fig. 4. See animation 3 for the all NVST and AIA images.
\label{fig2}}
\end{figure*}

\begin{figure*}[thbp]
\epsscale{0.9}
\plotone{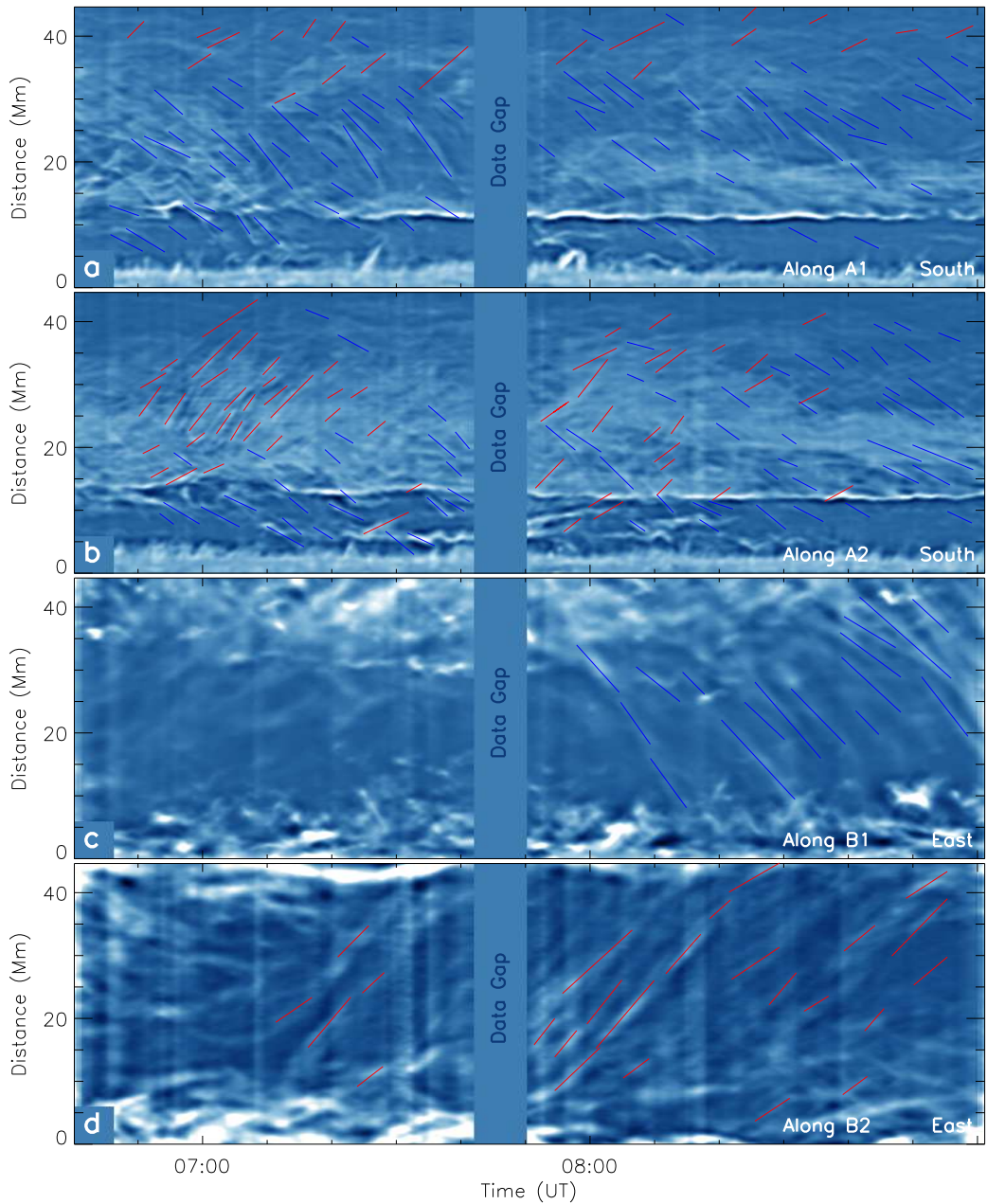}
\caption{The counter-streaming mass flows in vertical and horizontal feet of the prominence. a--b, time-distance diagrams along A1 and A2, and the red and blue lines mark the upward and downward mass flows, respectively. c--d, time-distance diagrams along B1 and B2, and the red and blue lines mark the westward and eastward mass flows, respectively. All these time-distance diagrams are made from the NVST H$\alpha$ data, and the data gap is indicated in each panel.
\label{fig3}}
\end{figure*}

\begin{figure*}[thbp]
\epsscale{0.85}
\plotone{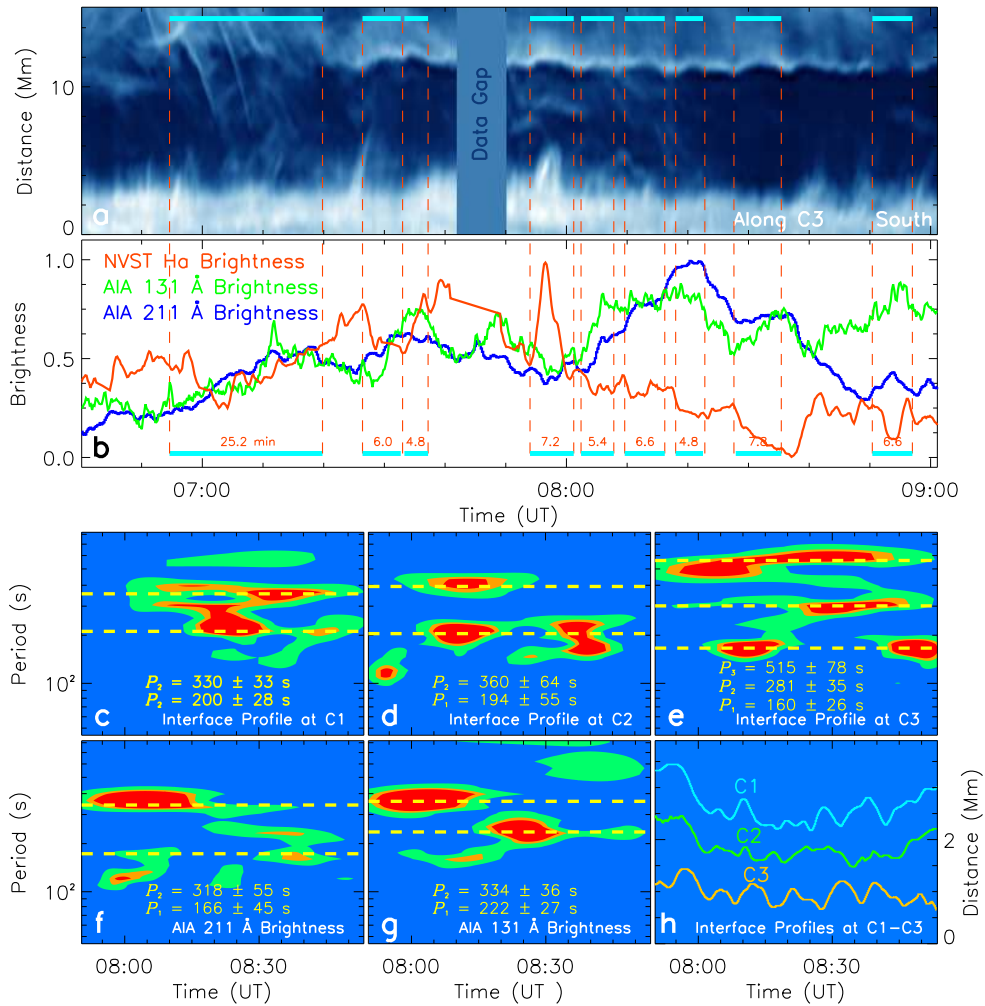}
\caption{The relationships between the dynamical bubble boundary and the brightness of the flare-like feature. a, time-distance diagram obtained from the NVST H$\alpha$ data along C3 as shown in Fig. 1. b, AIA 131 \AA\, 211 \AA, and NVST H$\alpha$ lightcurves show the brightness variations of the  flare-like feature. The vertical dashed lines and the cyan horizontal bars mark the duration of each oscillation motion. c--g, wavelet analysis of the oscillation trajectories of the bubble boundary at the positions of C1--C3 (c--e), the AIA 211 \AA\ (f) and 131 \AA\ (g) brightness profiles over the flare-like feature. h, the oscillation trajectories determined from the time-distance diagrams along C1--C3.
\label{fig4}}
\end{figure*}

\begin{figure*}[thbp]
\epsscale{0.85}
\plotone{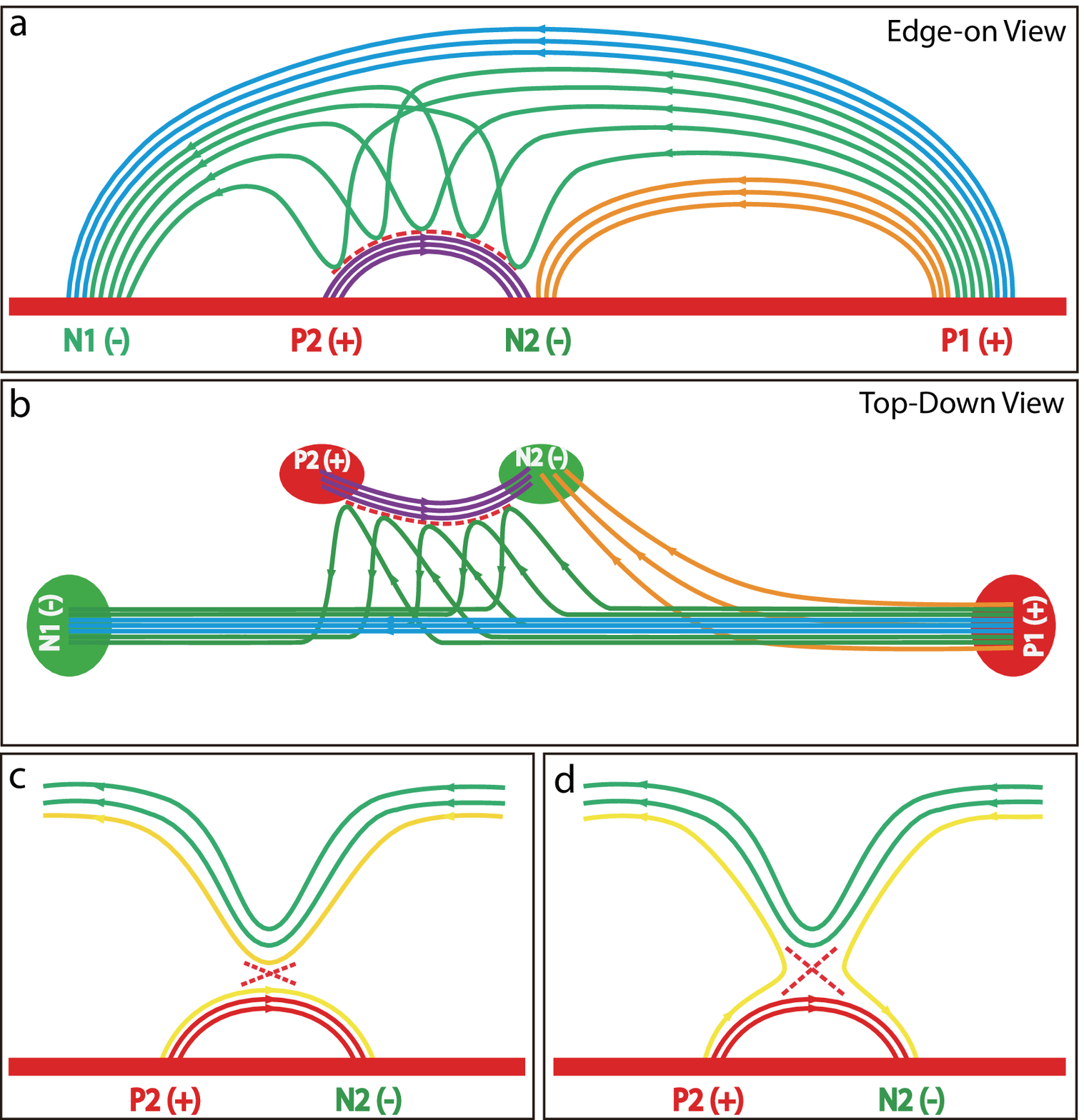}
\caption{A schematic model demonstrates the prominence magnetic structure. a, Edge-on view of the prominence. b, Top-down view of prominence. c, magnetic field configuration before the magnetic reconnection between the bubble and the overlying dips. d, magnetic configuration during the reconnection process. In panels a and b, the blue, green, orange, indigo, and red dashed lines represent the prominence spine, the vertical foot, the horizontal foot, the bubble, and the separator respectively. In panels c and d, the current sheet is marked by a dashed red cross. The yellow lines in panel c indicate the magnetic field lines to be reconnect, while those in panel d are the reconnected magnetic field lines. The horizontal thick red bar in panels a, c, and d represent the solar limb. Note that only some representative field lines are plotted in the figure.
\label{fig5}}
\end{figure*}

\section{RESULTS}
On 24 May 2014, a high-latitude quiescent prominence at the northwest solar limb ($28^{\circ}$ N, $88^{\circ}$ W) was observed by ground-based and space-borne telescopes such as NVST, KSO, and SDO. These instruments image the prominence at chromosphere and corona temperature simultaneously. The prominence was consisted of a horizontal and a vertical low-lying foot encircled by an overlying horizontal spine, and all of these structures are consisted of thin threads that manifest the magnetic field distribution in the prominence. While the horizontal foot seemingly with its both ends directly rooted on the solar surface, the vertical foot was suspended above the solar limb by a semicircular bubble structure (Fig. 1 a). The average width of the thin threads is about 700 km, and the height and width of the bubble are $13,000 \pm 336$ km and $25,000 \pm 576$ km, respectively. The AIA 171 \AA\ images show that the vertical foot was embedded in a large-scale loop structure, along which one can identify the prominence more clearly in AIA 304 \AA\ images (Fig. 1 b and c). The KSO H$\alpha$ telescope recorded the prominence since a few days ago. On 20 May 2014, the prominence was on the solar disk and showed up as a long filament consisting of a few discrete segments, and the one containing the bubble is indicated by the white arrows in Fig. 1 d--e. The HMI line-of-sight (LOS) magnetogram on 20 May indicates that the prominence was located on the polarity inversion line of a quiet-Sun region, and many minority polarities can be identified around the polarity inversion line. (Fig1. g).

The bubble below the vertical foot showed intriguing dynamical behaviours in the NVST H$\alpha$ images (Fig. 2 a--c). Firstly, it collapsed at about 06:55 UT and launched conspicuous upflows along the vertical threads. Since downflows were persistent along the vertical threads during the entire observing time, the combination of upflows and downflows formed a pattern of counter-streaming flows in the vertical foot. The collapsing also resulted in the successful intrusion of downflows into the bubble. Secondly, after the collapsing the bubble reformed at a similar height and then started to oscillate for about 90 minutes. Thirdly, during the oscillation stage the vertical threads above the bubble showed collective horizontal motion to the west at a speed of $4.1 \pm 1.2$ km s$^{-1}$. At the bottom of the bubble, an obvious flare-like feature is observed just above the solar surface in AIA EUV channels (here, only show AIA 211 \AA\ as an example), and the brightest stage occurred during the bubble's collapsing stage (Fig. 2 d--f). Using the differential emission measure (DEM) technique\citep[]{cheng12}, the DEM curves crossing the bubble and the prominence are derived respectively (Fig. 2 g). The result shows that the bubble has a similar temperature distribution to the prominence ($5.8 \leqslant {\rm log} T \geqslant 7.4$), and both the average temperature and the emission measure of the bubble are slightly higher than those of the prominence. It should be noted that since the DEM results are seriously influenced by the foreground and background off-limb emissions along the LOS\citep[]{parenti12},  the temperature of the prominence (bubble) may just represents the prominence-corona transition region (PCTR) (foreground and background off-limb emissions). Therefore, we can not tell whether the temperature of the bubble is hotter that the prominence or not.

Ubiquitous counter-streaming mass flows were observed in different parts of the prominence. In the vertical foot, downflows were persistent during the entire observing time, while upflows were heavily only during the bubble's collapsing stage, after which they weakened gradually and finally disappeared after 08:40 UT (Fig. 3 a and b). It is noted that the counter-streaming flows along the vertical threads are close to uniform motion, implying that the downward gravity must be balanced by some kind of upward forces such as thermal and magnetic pressure. The average speeds of the upward and downward flows are $13.7 \pm 5.2$ km s$^{-1}$ and $12.5 \pm 5.2$ km s$^{-1}$, respectively. In the horizontal foot (Fig. 3 c and d), the direction of the counter-streaming flows were eastward and westward for the lower and higher threads, and their average speeds are $20.1 \pm 4.5$ km s$^{-1}$ and $15.8 \pm 3.8$ km s$^{-1}$, respectively. The different speeds of the counter-streaming flows in the vertical and horizontal feet may indicate the different driving mechanisms of these counter-streaming mass flows. By comparing with previous studies, the speeds of counter-streaming flows are consistent with \cite{labrosse10} and \cite{schmieder10}, but they are less than the values reported by \cite{berger10}. However, these values are all of the same order.

By comparing the profiles of the oscillating bubble boundary and the brightness of the flare-like feature, one can find that the oscillation behaviour of the bubble boundary is consistent with the variation pattern of the EUV brightness of the flare-like feature (Fig. 4 a and b). Namely, the continuous large rising of the brightness of the flare-like feature corresponds to the collapsing stage of the bubble boundary, while those small rising of the brightness only led to moderate oscillation motions whose amplitude was about 250 km on average. This result may indicate that the energy released by the flare-like feature heated the bubble and therefore led to the vacation of the thermal and magnetic pressure inside the bubble, which would cause the collapsing and oscillation of the bubble boundary. It seems that the collapsing and oscillation of the bubble boundary are depending on the amplitude of the disturbance resulted from the flare-like feature. It should be noted that some sudden increase of the NVST H$\alpha$ brightness are caused by spicules. The collapsing of the bubble lasted for about 25.2 minutes, while the durations of the oscillating motions range from 4.8 to 7.8 minutes. Wavelet analysis\citep{torrence98} of the oscillating bubble boundary and the brightness of the flare-like feature profiles indicates that all these parameters contain two main periods of about three and five minute. \cite{schmieder13} also reported the period of five minute in a prominence, and they proposed that such oscillations are possibly associated with magnetic reconnection events in or just above the photosphere, or the leaking of photospheric characteristic oscillations into the chromosphere.

\section{INTERPRETATION AND DISCUSSION}
Based on the observation, we propose a cartoon model to interpret the magnetic structure of the prominence (Fig. 5 a and b), which is similar to the magnetohydrodynamics (MHD) model proposed by \cite{dudik12}. According to the model, the prominence is composed of a high-lying horizontal spine that directly connects to the solar surface, which encircles a low-lying horizontal and a vertical foot. The horizontal foot is composed of shorter field lines running partially along the spine and with the both ends connecting to the solar surface, while the vertical foot is consisted of piling-up dips due to the sagging of the initially horizontal spine fields. Meanwhile, the piling-up dips are supported by a low-lying closed magnetic system that forms the bubble structure. This topology includes a separator between the bubble and the overlying dips (red dashed curve in panels a and b of Fig. 5). If disturbance occurs around the separator, magnetic reconnection at the separator can be expect \citep{dudik12}. It should be point out that our interpretation of the vertical feet is consistent with the traditional theory for interpreting the vertical structures observed in solar prominences\citep[e.g.,][]{aulanier98,low05,chae08,schmieder10}. Using this model, one can naturally explain the coexistence of horizontal and vertical structures in a single prominence, the relationship between photospheric magnetic fields and the ends of prominence feet, and the origin of counter-streaming mass flows in prominences. Obviously, when one observes the prominence from the side, the spine and the horizontal foot will be observed as horizontal structures, while the vertical foot will be observed as vertical structure in prominence. When one observes the prominence from the top, the end of the vertical foot would be terminated at the small-scale polarity inversion line that forms by small polarities, and the ends of the horizontal foot are possibly rooted at small polarities directly (Fig. 5 b).

As pointed out by \citep{dudik12}, the sharp bubble boundary could either be due to the magnetic reconnection at the separator or due to the piling up of plasma near the separator, and the magnetic reconnection at the separator can be the driving mechanism of upflows (plumes) in prominences. For the present case, since the continuous heating of the flare-like feature, the thermal and magnetic pressure will rise continuously inside the bubble. This will disturb the initial equilibrium of the bubble and thus lead to the rise (or expansion) of the bubble. Therefore, the collapsing of the bubble can be interpreted as the result of the magnetic reconnection at the separator between the low-lying bubble and the overlying dips (Fig. 5 c and d), due to the large rising of temperature inside the bubble. The consequence of the reconnection are the appearance of the upflows in the vertical foot, the downflows intruded into the bubble, and the collective horizontal motion of the thin threads. If the disturbance results from the flare-like feature is too small to trigger the fast magnetic reconnection at the separator, the consequence may just be oscillation motions of the bubble boundary as we observed in the present case. The magnetic reconnection at the separator could also be a possible driving mechanism of counter-streaming mass flows in prominences. For the counter-streaming flows in the horizontal foot, we propose that they may caused by the imbalanced pressure at the both ends. Since the mass flows were apparently along different threads, the mechanism of longitudinal oscillation threads can be excluded. In this line of thought, counter-streaming flows can also be expected in the high-lying prominence spine, even though we do not detect counter-streaming mass flows in the spine based on the NVST H$\alpha$ observations.

Bubble structure is always observed in quiescent prominences, and it is possibly a fundamental building block of solar quiescent prominences. Earlier low-resolution ground-based\citep{stellmacher73,detoma08} and recent high-resolution space-borne\citep{berger08,berger10,berger11,dudik12} telescopes have observed several prominence bubbles and the generation of plumes that directly intrude into the prominence body. So far, the origin of prominence bubbles is still unclear. Berger et al.\citep{berger11} proposed that they are formed by emerging magnetic flux below prominences, but Gun\'{a}r et al.\citep{gunar14} suggested that they are formed because of perturbations of the prominence magnetic field by parasitic bipoles. In addition, the driving mechanism of prominence plumes from bubble structures is also unclear. Some authors proposed that the generation of plumes are caused by Rayleigh-Taylor instability at the bubble boundary\citep{berger11,hillier11,hillier12,keppens15}, while others suggest that they are possibly caused by magnetic reconnection at the boundary between the bubble and the prominence\citep{liu04,dudik12}. Our observations suggest that the prominence bubble is possibly formed by low-lying bipolar magnetic systems due to the appearance of small polarities near the photospheric polarity inversion line, and the generation of the upflows (or plumes) in the vertical foot are possibly drived by the magnetic reconnection at separator between the bubble and the overlying dips that forms the vertical prominence foot. However, the mechanism of Rayleigh-Taylor instability can not be completely discarded, and it may as well be present and play a role in the collapsing of the bubble and the generation of the upflows. The bubble's temperature and dynamics are inevitably influenced by the flare-like feature at the bottom of the bubble, which releases energy and changes the thermal and magnetic pressure inside the bubble and further disturbs the bubble boundary. The periodicity analysis of the flare-like feature's brightness and the bubble's boundary may indicate that the photospheric pressure-driven oscillations can effectively modulate the kinematics of prominence bubbles and the overlying prominence, via modulating the magnetic reconnection process that releases magnetic energy and thereby produced the flare-like feature \citep{chen06}..

\acknowledgments We thank the observations provided by the NVST, KSO, and {\em SDO} and an anonymous referee for many constructive suggestions. The NVST H$\alpha$ data  was from a joint campaign launched by Prof. B. Schmieder. This work is supported by the Natural Science Foundation of China (11403097), the Yunnan Science Foundation (2015FB191), the National Basic Research Program of China (973 program, 2011CB811400), the Specialized Research Fund for State Key Laboratories, the Open Research Program of the Key Laboratory of Solar Activity of Chinese Academy of Sciences (KLSA201401), the Key Laboratory of Modern Astronomy and Astrophysics of Nanjing University, the Youth Innovation Promotion Association (2014047) and the Western Light Youth Project of Chinese Academy of Sciences.



\newpage
\begin{thebibliography}{33}
\expandafter\ifx\csname natexlab\endcsname\relax\def\natexlab#1{#1}\fi
\bibitem[Ahn et al. (2010)]{ahn10}
Ahn, K., Chae, J., Cao, W., \& Goode, P., 2010, \apj, 721, 74
\bibitem[Amari et al. (2003)]{amari03}
Amari, T., Luciani, J. F., Aly, J. J., Miki\'{c}, Z., \& Linker, J., 2003, \apj, 595, 1231
\bibitem[Aulanier \& D\'{e}moulin (1998)]{aulanier98}
Aulanier, G., \& D\'{e}moulin, P., 1998, \aap, 329, 1125
\bibitem[Aulanier et al.(1998)]{aulanier98b}
Aulanier, G., \& D\'{e}moulin, P., van Driel-Gesztelyi, L., Mein, P., \& DeForest, C., 1998, \aap, 335, 309
\bibitem[Aulanier et al. (2002)]{aulanier02}
Aulanier, G., DeVore, C. R., \& Antiochos, S. K., 2002, \apj, 567, L97
\bibitem[Berger et al. (2008)]{berger08}
Berger, T. E., Shine, R. A., Slater, G. L., et al., 2008, \apj, 676, L89
\bibitem[Berger et al. (2010)]{berger10}
Berger, T. E., Slater, G., Hurlburt, N., et al., 2010, \apj, 716, 1288
\bibitem[Berger et al. (2011)]{berger11}
Berger, T. E., Testa, P., Hillier, A., et al., 2011, \nat, 472, 197
\bibitem[Casini et al. (2003)]{casini03}
Casini, R., L\'{o}pez Ariste, A., Tomczyk, S., \& Lites, B. W., 2003, \apj, 598, L67
\bibitem[Chae et al. (2008)]{chae08}
Chae, J., Ahn, K., Lim, E.-K., Choe, G. S., \& Sakurai, T., 2008, \apj, 689, L73
\bibitem[Chae et al. (2005)]{chae05}
Chae, J., Moon, Y. J., \& Park, Y. D., 2005, \apj, 626, 574
\bibitem[Chen et al. (2014)]{chen14}
Chen, P. F., Harra, L. K., \& Fang, C., 2014, \apj, 784, 50
\bibitem[Chen \& Priest (2006)]{chen06}
Chen, P. F., \& Priest, E. R., 2006, \solphys, 238, 313
\bibitem[Cheng et al. (2012)]{cheng12}
Cheng, X., Zhang, J., Saar, S. H., \& Ding, M. D., 2012, \apj, 761, 62
\bibitem[de Toma et al. (2008)]{detoma08}
de Toma, G., Casini, R., Burkepile, J. T., \& Low, B. C., 2008, \apj, 687, L123
\bibitem[Dud\'{i}k et al. (2012)]{dudik12}
Dud\'{i}k, J., Aulanier, G., Schmieder, B., Zapi\'{o}r, M., \& Heinzel, P., 2012, \apj, 761, 9
\bibitem[Gun\'{a}r et al. (2014)]{gunar14}
Gun\'{a}r, S., Schwartz, P., Dud\'{i}k, J., et al., 2014, \aap, 567, A123
\bibitem[Heinzel et al. (2008)]{heinzel08}
Heinzel, P., Schmieder, B., F\'{a}rn\'{i}k, F., Schwartz, P., Labrosse, N., et al., 2008, \apj, 686, 1383
\bibitem[Hillier et al. (2012)]{hillier12}
Hillier, A., Berger, T., Isobe, H., \& Shibata, K., 2012, \apj, 746, 120
\bibitem[Hillier et al. (2011)]{hillier11}
Hillier, A., Isobe, H., Shibata, K., \& Berger, T., 2011, \apj, 736, L1
\bibitem[Keppens et al.(2015)]{keppens15}
Keppens, R., Xia, C., \& Porth, O., 2015, \apj, 806, L13
\bibitem[Kippenhahn et al. (1957)]{kippenhahn57} 
Kippenhahn, R., \& Schl\"{u}term, A.1957, {\it Z. Astrophys.}, 43, 36
\bibitem[Kuperus \& Raadu (1974)]{kuperus74}
Kuperus, M., \& Raadu, M. A., 1974, \aap, 31, 189
\bibitem[Labrosse et al.(2010)]{labrosse10}
Lacrosse, N., Heinzel, P., Vial, J.-C., et al., 2010, \solphys, 151, 243
\bibitem[Lemen et al. (2012)]{lemen12}
Lemen, J. R.,Title, A. M., Akin, D. J., et al., 2012, \solphys, 275, 17
\bibitem[Li \& Zhang (2013)]{li13}
Li, L. P., \& Zhang, J., 2013, \solphys, 282, 147
\bibitem[Lin et al. (2005)]{lin05}
Lin, Y., Engvold, O., Rouppe van der Voort, L., Wiik, J. E., \& Berger, T. E., 2005, \solphys, 226, 239
\bibitem[Lin et al. (2003)]{lin03}
Lin, Y., Engvold, O., \& Wiik, J. E., 2003, \solphys, 216, 109
\bibitem[Liu \& Kurokawa (2004)]{liu04}
Liu, Y., \& Kurokawa, H., 2004, \pasj, 56, 497
\bibitem[Liu et al. (2014)]{liuy14}
Liu, Y. D., Luhmann, J. G., Kajdi\v{c}, P., et al., 2014, NatCo, 5, 3481
\bibitem[Liu et al. (2014)]{liuz14}
Liu, Z., Xu, J., Gu, B. Z., et al., 2014, RAA,14, 705
\bibitem[Low \& Petrie (2005)]{low05}
Low, B. C., \& Petrie, G. J. D., 2005, \apj, 626, 551
\bibitem[Mackay et al.(2010)]{mackay10}
Mackay, D. H., Karpen, J. T., Ballester, J. L., Schmieder, B., \& Aulanier, G., 2010, \ssr, 151, 333
\bibitem[Martin (1998)]{martin98}
Martin, S. F., 1998, \solphys, 182, 107
\bibitem[Martin \& Echols (1994)]{martin94}
Martin, S.F., \& Echols, C.R.: 1994, In: Rutten, R.J., Schrijver, C.J. (eds.) Solar Surface Magnetism, 339
\bibitem[Okamoto et al. (2007)]{okamoto07}
Okamoto, T. J., Tsuneta, S., Berger, T. E., et al., 2007, Science, 318, 1577
\bibitem[Parenti(2012)]{parenti12}
Parenti, S., Schmieder, B., Heinzel, P., \& Golub, L., 2012, \apj, 754, 66
\bibitem[Pesnell et al. (2012)]{pesnell12}
Pesnell, W. D, Thompson, B. J., \& Chamberlin, P. C., 2012, \solphys, 275, 3
\bibitem[Schmieder et al.(2008)]{schmieder08}
Schmieder, B., Bommier, V., Kitai, R., et al., 2008, \solphys, 247, 321
\bibitem[Schmieder et al. (2010)]{schmieder10}
Schmieder, B., Chandra, R., Berlicki, A., \& Mein, P., 2010, \aap, 514, A68
\bibitem[Schmieder et al.(2013)]{schmieder13}
Schmieder, B., Kucera, T. A., Knizhnik, K., et al., 2013, \apj, 777, 108
\bibitem[Schmieder et al.(2015)]{schmieder15}
Schmieder, B., Aulanier, G., Vr\v{s}nak, B., 2015, \solphys, tmp, 64 
\bibitem[Schou et al. (2012)]{schou12}
Schou, J., Borrero, J. M., Norton, A. A., et al., 2012, \solphys, 275, 229
\bibitem[Shibata \& Magara (2011)]{shibata11}
Shibata, K., \& Magara, T., 2011, LRSP, 8, 6
\bibitem[Stellmacher \& Wiehr (1973)]{stellmacher73}
Stellmacher, G., \& Wiehr, E., 1973, \aap, 24, 321
\bibitem[Tandberg-Hanssen (1995)]{tandberg95}
Tandberg-Hanssen, E. (ed.) 1995, The Nature of Solar Prominences, Vol. 199 (Dordrecht: Kluwer Academic)
\bibitem[Torrence \& Compo (1998)]{torrence98}
Torrence, C., \& Compo, G. P., 1998, {\it Bull. Meteorol. Soc.}, 79, 61
\bibitem[Wang (2001)]{wang01}
Wang, Y. M., 2001, \apj, 560, 456
\bibitem[Yan et al. (2015)]{yan15}
Yan, X. L., Xue, Z. K., Xiang, Y. Y., \& Yang, L. H., 2015, 10,1725
\bibitem[Yang et al. (2014)]{yang14}
Yang, S., Zhang, J., Liu, Z., \& Xiang, Y., 2014, \apj, 784, L36
\bibitem[Zirker et al. (1998)]{zirker98}
Zirker, J. B., Engvold, O., \& Martin, S. F., 1998, \nat, 396, 440
\end{thebibliography}
\end{document}